\documentclass[epj]{svjour}
\usepackage{graphicx}
\usepackage{amssymb,amsbsy,amsmath}
\newcommand{\op}[1]{%
    \fontdimen12\textfont3=2pt\fontdimen12\scriptfont3=1.4pt%
    \!\null\mathop{\vphantom{#1}\smash{#1}}\limits_{\sim}\null\!}
\newcommand{\xref}[1]{\protect\ref{#1}}
\newcommand{\figref}[1]{Fig.~\protect\ref{#1}}
\newcommand{\fmref}[1]{(\protect\ref{#1})}

\newcommand {\telcomp}{Sr$_{14}$Cu$_{24}$O$_{41}$}
\journalname{Eur. Phys. J. B}
\begin{document}
\title{Strong Coulomb effects in hole-doped Heisenberg chains}
\author{J\"urgen Schnack%
%\inst{1}
}                     % Do not remove
\offprints{J\"urgen Schnack}          % Insert a name or remove this line
\institute{Universit\"at Osnabr\"uck, Fachbereich Physik,
D-49069 Osnabr\"uck, Germany}
\date{Received: date / Revised version: date}
\abstract{
  Substances such as the ``telephone number compound'' \telcomp\
  are intrinsically hole-doped. The involved interplay of spin
  and charge dynamics is a challenge for theory. In this article
  we propose to describe hole-doped Heisenberg spin rings by
  means of complete numerical diagonalization of a Heisenberg
  Hamiltonian that depends parametrically on hole positions and
  includes the screened Coulomb interaction among the holes.  It
  is demonstrated that key observables like magnetic
  susceptibility, specific heat, and inelastic neutron
  scattering cross section depend sensitively on the dielectric
  constant of the screened Coulomb potential.
 % end of PACS codes
} %end of abstract
\maketitle
%

%%%%%%%%%%%%%%%%%%%%%%%%%%%%%%%%%%%%%%%%%%%%%%%%%%%%%%%%%%%%%%%%%%%%%%%%
\section{Introduction and model}
\label{sec-1}

Substances hosting spin and charge degrees of freedom exhibit a
large variety of phenomena like magnetic and charge ordering,
metallic conductivity and superconductivity
\cite{Dag:RMP94,KCG:PRB01}. The ``telephone number compound'',
\linebreak \telcomp , contains two magnetic one-dimensional
structures, chains and ladders.  The stoichiometric formula
suggests 6 holes per formula unit. We will assume that for the
undoped compound all holes are located in the chain subsystem
(i.~e. 60~\% holes), although this is experimentally under
discussion since x-ray absorption (XAS) measurements suggest
that at room temperature some holes are located in the ladder
subsystem \cite{NMK:PRB00}, whereas it is necessary to assume
that all holes are in the chain subsystem in order to explain
neutron scattering data \cite{RBM:PRB99}.

At low temperatures the ladder subsystem is magnetically
inactive due to a large spin gap \cite{TME:PRB98}.  The
remaining dynamics of the hole-doped chain system is still
interesting as well as complicated enough to constitute a
challenge for theoretical investigations. Especially the
evaluation of thermodynamic quantities both as function of
temperature and magnetic field is prohibitively complicated even
for moderate system sizes. Therefore, mostly approximate
descriptions in terms of classical spin dynamics
\cite{HoS:PRE03,SPV:EPJB02}, spin-dimer models
\cite{CBC:PRL96,ABC:PRB00,Klingeler:2003} or spin-wave analysis
\cite{MYK:PRB99} have been applied.  Calculations based on the
Hubbard model aim at ground-state correlations at low hole
doping \cite{AKA:PRB98}.

\begin{figure}[ht!]
\centering
%\resizebox{0.75\textwidth}{!}{\includegraphics[clip,width=65mm]{fig-1.eps}}
\includegraphics[clip,width=65mm]{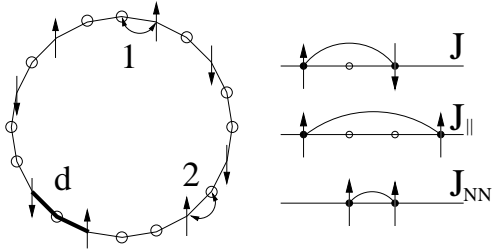}
\caption{L.h.s.: Ground-state hole configuration for 20 sites and 60\%
  holes. This configuration is also called dimer configuration
  since it consists of weekly interacting antiferromagnetic
  dimers. One dimer is highlighted (d). The hole-spin exchange
  processes \textbf{1} and \textbf{2} lead to energetically
  low-lying configurations.
  R.h.s.: Exchange parameters used in this article: $J=-67$~K,
  $J_\parallel=7$~K, and $J_{NN}=25$~K.
}
\label{F-1}
\end{figure}

A fundamental question in this context is how the charge order
in the CuO$_2$ chains of substances such as \telcomp\ is
established. These chains seem to be a sequence of rather
perfect antiferromagnetically coupled spin dimers separated by
holes, see \figref{F-1}.  Any proposed theoretical model should
also be able to describe excitations involving hole motion which
is crucial because interesting physical properties of these
compounds result from a competition of charge mobility and
magnetic interactions
\cite{KFE:Nature98,MYD:Science99,Zaanen:Science99,HEB:PRL04}.
One possible explanation is that the formation of dimers is
generated by structural modulations of the material via a strong
variation of the on-site orbital energies
\cite{IsT:JPSJ98,GeL:PRL04,GeL:EPJB05,GeL:05}.

In this article we investigate how a screened electrostatic
hole-hole repulsion along the chain would express itself in
thermodynamic quantities. It will turn out that a rather strong
Coulomb repulsion is needed in order to reproduce the
experimental magnetization. This is in accord with e.g.
Ref.~\cite{VFB:PRB03} or with
Refs.~\cite{MTM:PRB98A,MTM:PRB98B}, where a dielectric constant
of 3.3 is found to be realistic.

In order to be able to evaluate thermodynamic quantities we
propose to describe hole-doped spin rings with a Heisenberg
Hamiltonian that depends parametrically on hole positions.  This
ansatz is similar to a simple Born-Oppenheimer description where
the electronic Hamiltonian (here spin Hamiltonian) depends
parametrically on the positions of the classical nuclei (here
hole positions).  Each configuration $\vec{c}$ of holes and
spins defines a Hilbert space which is orthogonal to all Hilbert
spaces arising from different configurations. The Hamilton
operator $\op{H}(\vec{c})$ of a certain configuration $\vec{c}$
is of Heisenberg type and depends parametrically on the actual
configuration $\vec{c}$, i.~e.
%--------------------------------------------------------
\begin{eqnarray}
\label{E-1-1}
\op{H}
&=&
\sum_{\vec{c}}\;
\left(
\op{H}(\vec{c})
+
V(\vec{c})
\right)
\\
\label{E-1-2}
\op{H}(\vec{c})
&=&
-
\sum_{u,v}\;
J_{uv}(\vec{c})\;
\op{\vec{s}}(u) \cdot \op{\vec{s}}(v)
\ .
\end{eqnarray}
%--------------------------------------------------------
$J_{uv}(\vec{c})$ are the respective exchange parameters which
depend on the configuration of holes. $J<0$ describes
antiferromagnetic coupling, $J>0$ ferromagnetic coupling.

%===================    figure   =================================
%===================    figure   =================================

Figure~\xref{F-1} shows on the l.h.s. as an example the
ground-state hole configuration of \telcomp\
\cite{RBM:PRB99,MYK:PRB99,ABC:PRB00} which is a sequence of
spin-hole-spin dimers separated by two holes. Energetically
excited configurations arise if holes are moved to other sites
as depicted exemplarily by the exchange processes \textbf{1} and
\textbf{2}.  The r.h.s. of \figref{F-1} illustrates how the
exchange parameters depend on the actual hole configuration. In
this work three different exchange parameters are employed.

A key ingredient of the proposed model is the inclusion of the
electrostatic interaction between holes which is modeled by a
screened Coulomb potential
%--------------------------------------------------------
\begin{eqnarray}
\label{E-2}
V(\vec{c})
&=&
\frac{e^2}{4\pi\epsilon_0\,\epsilon_r\,r_0}
\frac{1}{2}\;
\sum_{u\ne v}\;
\frac{1}{|u-v|}
\ ,
\end{eqnarray}
%--------------------------------------------------------
where $r_0=2.75$~\AA\ is the distance between nearest neighbor
sites on the ring. The dielectric constant $\epsilon_r$ is
considered as the only free parameter in the present
investigation. Several attempts have been undertaken to estimate
the dielectric constant which yielded values for $\epsilon_r$ up
to 30 \cite{CPP:PRL89,BaS:PRB94,EKZ:PRB97}. In related projects
where the exchange interaction of chain systems in cuprates is
derived from hopping matrix elements between different orbitals
using a Madelung potential the dielectric constant is found to
be $\epsilon_r=3.3$ \cite{MTM:PRB98A,MTM:PRB98B}.

%%%%%%%%%%%%%%%%%%%%%%%%%%%%%%%%%%%%%%%%%%%%%%%%%%%%%%%%%%%%%%%%%%%%%%%%
\section{Discussion of the model}
\label{sec-2}

The aim of the proposed model is to evaluate the complete
spectrum for reasonably large system sizes and thus to be able
to investigate thermodynamic quantities both as function of
temperature and field. The spectrum does not only consist of
those levels belonging to the ground-state hole distribution,
compare \figref{F-1}, but also of all levels arising from all
other hole configurations.  For small systems all such
configurations can be generated and the related spin
Hamiltonians \fmref{E-1-2} can be diagonalized completely. For 8
spins and 12 holes for instance this amounts to 6310 distinct
hole configurations and tiny Hilbert spaces of dimension 256.
For 16 spins and 24 holes the total number of hole
configurations is already too big to be considered completely.
Therefore, only the ground-state configuration and low-lying
excitations with their respective degeneracies are taken into
account. This is sufficient since hole configurations which
deviate considerably from the ground state configuration possess
very high excitation energies.

Correlated electrons are usually modeled with the Hubbard model
\cite{AKA:PRB98}, therefore looking at Eqs.~\fmref{E-1-1} and
\fmref{E-1-2} one might be tempted to ask: \emph{Where is the
  kinetic energy of the holes?} The Hamiltonian of the Hubbard
model \cite{Hubbard63,Hubbard64a,Hubbard64b},
%--------------------------------------------------------
\begin{eqnarray}
\label{E-3}
\op{H}
&=&
- \sum_{\langle ij \rangle,\sigma} t_{ij} 
\left(
\op{c}^\dagger_{i\sigma}\op{c}_{j\sigma} + \text{h.c.} \right) 
+ U \sum_{i} \op{n}_{i\uparrow}\op{n}_{i\downarrow}
\ ,
\end{eqnarray}
%--------------------------------------------------------
transforms at large $U$ and half filling into a Heisenberg
Hamiltonian \fmref{E-1-2} with $J_{ij}=-4t_{ij}^2/U$. Therefore,
if working close to half filling, it is legitimate to say that
the kinetic energy is absorbed into the exchange coupling.
Nevertheless, the remaining hole motion is treated classically,
i.~e. superpositions of hole states are not taken into account.
For the actual compound which is far from superconductivity this
assumption of well-localized holes seems to be appropriate.

What is properly taken into account is the screened Coulomb
interaction between holes. But, if so: \emph{Wouldn't it be
  sufficient to consider nearest-neighbor Coulomb repulsion
  only?} Although this is a common method we find that a simple
nearest-neighbor repulsion results in unphysical ground states. 
It is experimentally verified by means of low-temperature
susceptibility \cite{Klingeler:2003}, neutron scattering
\cite{RBM:PRB99,MYK:PRB99} as well as thermal expansion
measurements \cite{ABC:PRB00}, that the highly symmetric dimer
configuration, compare \figref{F-1}, constitutes the ground
state of \telcomp. Using only nearest-neighbor Coulomb repulsion
yields an alternating sequence of spins and holes with the
remaining 10~\% holes assembling as a big cluster irrespective
how strong the repulsion is. The reason is that this strange
configuration has the same number of nearest hole-hole neighbors
as the dimer configuration.  Even the inclusion of a
next-nearest neighbor Coulomb repulsion does not improve the
situation, the Coulomb interaction is still proportional to the
number of sites and may be overcome by the antiferromagnetic
binding $J$.

%%%%%%%%%%%%%%%%%%%%%%%%%%%%%%%%%%%%%%%%%%%%%%%%%%%%%%%%%%%%%%%%%%%%%%%%
\section{Results}
\label{sec-3}

At temperatures and energies below 200~K the behavior of the
chain subsystem in \telcomp\ is usually discussed in terms of
weakly interacting dimers sometimes augmented by weak interchain
interactions, see e.~g. 
\cite{CBC:PRL96,MYK:PRB99,ABC:PRB00,Klingeler:2003}. Such a
picture, although rather successful, does not allow to discuss
the influence of mobile holes on thermodynamic observables. It
is clear that configurations like \textbf{1} and \textbf{2} in
\figref{F-1} will contribute to thermal averages, but how? 

%===================    figure   =================================
\begin{figure}[ht!] 
\centering
%\resizebox{0.75\textwidth}{!}{\includegraphics{fig-2.eps}}
\includegraphics[clip,width=60mm]{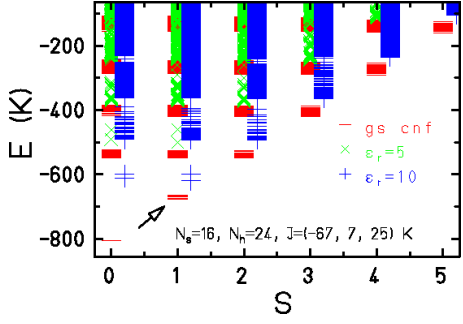}
\caption{Low-energy part of the spectrum of a chain of
  $N_{\text{tot}}=40$ sites with $N_{\text{s}}=16$ spins and
  $N_{\text{h}}=24$ holes for three choices of the dielectric
  constant $\epsilon_r$: bars -- $\epsilon_r=1$, bars together
  with x-symbols -- $\epsilon_r=5$, bars and crosses --
  $\epsilon_r=10$.  The arrow marks the singlet-triplet
  transition employed in dimer models.} 
\label{F-2}
\end{figure}
%===================    figure   =================================

Figure \xref{F-2} shows the low-energy part of the spectrum of a
chain of $N_{\text{tot}}=40$ sites with $N_{\text{s}}=16$ spins
and $N_{\text{h}}=24$ holes for three choices of the dielectric
constant $\epsilon_r$.  If $\epsilon_r=1$ the spectrum up to
several hundreds of Kelvin is solely given by the levels of the
ground-state dimer configuration (bars in \figref{F-2}).  With
increasing $\epsilon_r$ the Coulomb repulsion decreases and so
does the excitation energy of magnetic levels belonging to hole
configurations where one or two holes are moved. As an example
the levels resulting from such configurations are given as
x-symbols ($\epsilon_r=5$) and crosses ($\epsilon_r=10$) in
\figref{F-2}. It is clear that besides the singlet-triplet
transition (arrow in \figref{F-2}), which is the main ingredient
of the dimer model, transitions to states involving spin-holes
exchange processes will contribute to thermodynamic observables
like the inelastic neutron scattering cross section. In the
following we discuss the influence on three basic observables. 

%===================    figure   =================================
\begin{figure}[ht!] 
\centering
%\resizebox{0.75\textwidth}{!}{\includegraphics{fig-3.eps}}
\includegraphics[clip,width=60mm]{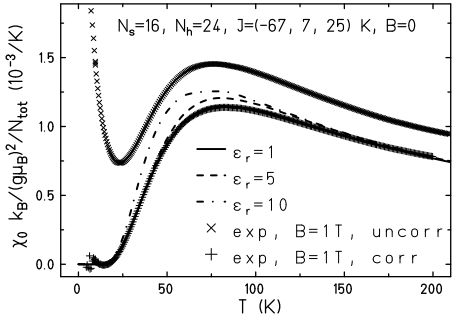}
\caption{Zero-field magnetic susceptibility $\chi_0(T)$: solid
  curve -- $\epsilon_r=1$, dashed curve -- $\epsilon_r=5$,
  dashed-dotted curve -- $\epsilon_r=10$. For a comparison
  experimental data, taken at $B=1$~T, are provided by x-symbols
  \cite{Klingeler:2003}. The corrected data, given by crosses,
  take also into account that due to impurities the number of
  dimers is less than theoretically possible
  \cite{Klingeler:2003}.}
\label{F-3}
\end{figure}
%===================    figure   =================================

Figure~\xref{F-3} presents the results for the magnetic
susceptibility $\chi(T,B=0)=\chi_0(T)$ at vanishing magnetic
field $B=0$. The solid curve shows the theoretical
susceptibility for $\epsilon_r=1$, the dashed curve for
$\epsilon_r=5$, and the dashed-dotted curve for $\epsilon_r=10$.
One realizes that with increasing $\epsilon_r$, i.~e. with
stronger screening of the Coulomb interaction, the
susceptibility increases at intermediate temperatures and that
the maximum shifts to lower temperatures. Although being rather
moderate it is astonishing that the effect is at all observable
since the responsible levels are at excitation energies well
above that temperature range, compare the spectrum in
\figref{F-2}.

It turns out that the high degeneracy of excited hole
configurations is the reason for the influence of the hole
dynamics even at low temperatures. Looking again at the hole
configuration shown in \figref{F-1} one notices that the
spin-hole exchange processes can happen at very different places
leading to a large geometric degeneracy. This degeneracy can
overcompensate a small Boltzmann factor and thus expresses
itself in a high thermal weight. 

Together with the theoretical results \figref{F-3} shows the
experimentally determined magnetization which was measured along
the $c$-axis of the material at a magnetic field of $B=1$~T
\cite{Klingeler:2003}. The x-symbols depict the uncorrected
values whereas the crosses represent the corrected values. The
correction includes a subtraction of impurities (free spins
$s=1/2$) as well as a rescaling because the number of dimers on
the chain is less than theoretically possible. The almost
perfect coincidence with the theoretical result for
$\epsilon_r=1$ suggests that the hole-hole repulsion is rather
strong. The uncertainties in the measurement and the correction
procedure leave some freedom for the actual value, but it is
clear that $\epsilon_r$ should not be bigger than about three.
This implies that energy levels which result from other than the
dimer configuration are well above the triplet excitation,
compare \figref{F-2}. Although this might seem to be unrealistic
one has to keep in mind that any theoretical model must explain
why the experimental susceptibility practically coincides with
that of free dimers. This is only possible if other excitations
are well separated from the triplet excitation.

%===================    figure   =================================
\begin{figure}[ht!] 
\centering
%\resizebox{0.75\textwidth}{!}{\includegraphics{fig-4.eps}}
\includegraphics[clip,width=60mm]{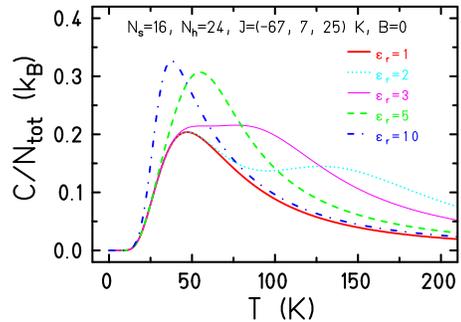}
\caption{Specific heat at $B=0$: $\epsilon_r=1$ -- solid curve,
  $\epsilon_r=2$ -- dotted curve, $\epsilon_r=3$ -- thin curve,
  $\epsilon_r=5$ -- dashed curve, and $\epsilon_r=10$ --
  dashed-dotted curve.} 
\label{F-4}
\end{figure}
%===================    figure   =================================

But even if higher-lying energy levels are well above the
triplet excitation, due to their vast degeneracy they can
substantially contribute to the specific heat, which is shown in
\figref{F-4}. In order to demonstrate how the thermal weight of
the excited hole configurations grows several cases are shown.
The solid curve, which is the lowest among all curves, again
depicts the result for $\epsilon_r=1$.  With increasing
dielectric constants thermal weight is shifted from higher
temperatures to lower ones. This can very clearly be seen for
the shown sequence: $\epsilon_r=2$ (dotted curve),
$\epsilon_r=3$ (thin curve), $\epsilon_r=5$ (dashed curve), and
$\epsilon_r=10$ (dashed-dotted curve). This result shows that
for dielectric constants of the order of $\epsilon_r\approx 5$
about one third of the specific heat at its maximum is due to
states involving hole motion.

Inelastic neutron scattering is a valuable tool to measure the
magnetic excitation spectrum of substances like \telcomp, see
e.~g. \cite{RBM:PRB99,BRL:PB04}. The transition from the singlet
ground state to the first excited triplet state, arrow in
\figref{F-2}, has been measured with high accuracy and used to
determine the exchange constants, especially $J$ (\figref{F-1},
r.h.s.). For recent results have a look at \cite{BRL:PB04}. 

%===================    figure   =================================
\begin{figure}[ht!] 
\centering
%\resizebox{0.75\textwidth}{!}{\includegraphics{fig-5.eps}}
\includegraphics[clip,width=60mm]{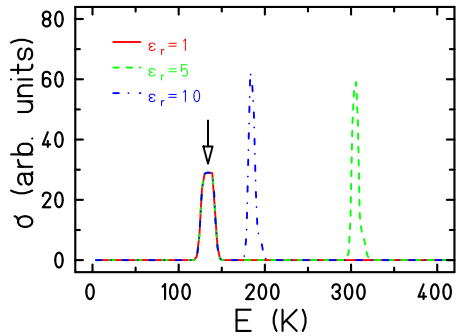}
\caption{Rough sketch of the lowest transitions observable with
  inelastic neutron scattering: $\epsilon_r=1$ -- solid curve,
  $\epsilon_r=5$ -- dashed curve, and $\epsilon_r=10$ --
  dashed-dotted curve. The arrow marks the singlet-triplet
  transition at about 135~K.} 
\label{F-5}
\end{figure}
%===================    figure   =================================

In addition to this fundamental transition of the dimer
configuration, transitions to states with spin-hole exchange
should be detectable, too. Figure~\xref{F-5} shows as a rough
sketch where such transitions could be expected. The
singlet-triplet transition at about 135~K -- arrow in
\figref{F-5} -- is clearly seen for each dielectric constant. 
But with increasing $\epsilon_r$ transitions to states with one
or two holes moved become accessible. The dashed curve shows
schematically the lowest transitions for $\epsilon_r=5$, the
dashed-dotted curve the lowest transitions if $\epsilon_r=10$. 
Although the singlet-triplet transition might have the largest
matrix element, again the huge geometric degeneracy could help
to make the other transitions visible. A possible drawback is,
nevertheless, due to the fact that states involving spin-hole
exchange break the translational symmetry, compare \figref{F-1}. 
The momentum dependency, therefore, might be fuzzy.

%%%%%%%%%%%%%%%%%%%%%%%%%%%%%%%%%%%%%%%%%%%%%%%%%%%%%%%%%%%%%%%%%%%%%%%%
\section{Summary and Outlook}
\label{sec-4}

For the compound under investigation the model
is capable to address much more questions. Since it is not clear
whether the hole content of the chain subsystem in \telcomp\ is
really 60~\%, one can study the influence of a reduced number of
holes on magnetic observables. 

%===================    figure   =================================
\begin{figure}[ht!] 
\centering
%\resizebox{0.75\textwidth}{!}{\includegraphics{fig-6.eps}}
\includegraphics[clip,width=60mm]{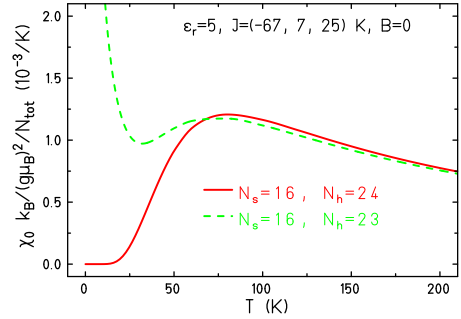}
\caption{Zero-field magnetic susceptibility $\chi_0(T)$ for
  $\epsilon_r=5$: solid curve -- $N_{\text{s}}=16$ and
  $N_{\text{h}}=24$, dashed curve  -- $N_{\text{s}}=16$ and
  $N_{\text{h}}=23$.} 
\label{F-6}
\end{figure}
%===================    figure   =================================

Figure~\xref{F-6} shows as an example how a reduction by just
one hole will influence the magnetic susceptibility. The perfect
symmetry of successive spin-hole-spin dimers separated by two
holes is destroyed and a low-lying triplet competes with the
former singlet ground state which leads to a low-temperature
divergence of the susceptibility. It might very well be that a
part of the experimentally observed low-temperature divergence
of the susceptibility \cite{Klingeler:2003} is due to the
reduced hole doping of the chain. 

The absence of perfect symmetry also leads to an increased
mobility of the holes. At the imperfections neighboring holes
can be moved without altering the Coulomb energy much. 
Therefore, the excitation energy for configurations where a hole
is moved at imperfections will be rather low.

Summarizing, the main advantage of the proposed effective
spin-hole Hamiltonian is that it allows to evaluate
thermodynamic observables both as function of temperature and
magnetic field for reasonably large systems. Using this model it
could be shown how a hole-hole Coulomb repulsion along the chain
would express itself in thermodynamic observables. The
comparison with experimental magnetization data suggests that
the screening is weak. The actual choice of the exchange
parameters, compare \figref{F-1}, does not influence the general
conclusions about the effects of the Coulomb repulsion between
holes.  It is not yet clear whether a weekly screened Coulomb
repulsion is realistic \cite{VFB:PRB03} or whether a combination
of hole-hole Coulomb repulsion and modulation of on-site
energies \cite{IsT:JPSJ98,GeL:PRL04,GeL:EPJB05,GeL:05} has to be
used. An experimental determination of other excited states than
the triplet excitation of the dimers would be very helpful in
this respect.

%%%%%%%%%%%%%%%%%%%%%%%%%%%%%%%%%%%%%%%%%%%%%%%%%%%%%%%%%%%%%%%%%%%%%%%%
\section*{Acknowledgement}

I would like to thank Bernd B\"uchner, R\"udiger Klingeler,
Fatiha Ouchni, and Louis-Pierre Regnault for fruitful
discussions, R\"udiger Klingeler for providing the experimental
magnetization data, and Heinz-J\"urgen Schmidt for carefully
reading the manuscript.

%\bibliographystyle{/usr/share/texmf/tex/latex/local/revtex/prsty}
%\bibliography{/home/schnack/tex/bibtex/js-own,/home/schnack/tex/bibtex/js-mag,/home/schnack/tex/bibtex/js-cup,/home/schnack/tex/bibtex/js-mis}

\end{document}